\documentclass[conference]{IEEEtran}
\usepackage{amsmath,amssymb,amsfonts}
\usepackage{algorithmic}
\usepackage{graphicx}
\usepackage{algorithm}
\usepackage{algorithmic}
\usepackage{textcomp}
\usepackage[usenames,dvipsnames]{xcolor}
\usepackage{listings}
\usepackage[keeplastbox,nospread]{flushend}
\usepackage[hidelinks,colorlinks,
            linkcolor={red!80!black},
            citecolor={green!80!black},
            urlcolor={blue!80!black}]{hyperref}
\usepackage{url}
\usepackage{booktabs}
\usepackage{multirow}
\usepackage{physics}
\usepackage{cite}
\def\BibTeX{{\rm B\kern-.05em{\sc i\kern-.025em b}\kern-.08em
T\kern-.1667em\lower.7ex\hbox{E}\kern-.125emX}}

\begin{document}

\title{One-Shot Template Matching for Automatic Document Data Capture}

\makeatletter
\newcommand{\linebreakand}{
  \end{@IEEEauthorhalign}
  \hfill\mbox{}\par
  \mbox{}\hfill\begin{@IEEEauthorhalign}
}
\makeatother

\author{
    \IEEEauthorblockN{Pranjal Dhakal, Manish Munikar and Bikram Dahal}
    \IEEEauthorblockA{\emph{Docsumo}\\
        \tt \{pranjal.dhakal, manish.munikar, bikram.dahal\}@docsumo.com
    }
}

% \author{
%   \IEEEauthorblockN{Pranjal Dhakal}
%   \IEEEauthorblockA{\textit{Docsumo}\\
%     \tt pranjal.dhakal@docsumo.com}
%   \and
%   \IEEEauthorblockN{Manish Munikar}
%   \IEEEauthorblockA{\textit{Docsumo}\\
%     \tt manish.munikar@docsumo.com}
%   \and
%   \IEEEauthorblockN{Bikram Dahal}
%   \IEEEauthorblockA{\textit{Docsumo}\\
%     \tt bikram.dahal@docsumo.com}
% }

\maketitle

\begin{abstract}
  In this paper, we propose a novel one-shot template-matching algorithm to automatically capture data from business documents with an aim to minimize manual data entry. Given one annotated document, our algorithm can automatically extract similar data from other documents having the same format. Based on a set of engineered visual and textual features, our method is invariant to changes in position and value. Experiments on a dataset of 595 real invoices demonstrate 86.4\% accuracy.
\end{abstract}

\begin{IEEEkeywords}
  document processing, automatic data capture, template matching, one-shot
  learning
\end{IEEEkeywords}
\section{Introduction}

Every business needs to process a lot of documents such as invoices, bills,
statements, forms, etc. saved in unstructured formats such as PDF or scanned
images into their accounting software. The larger ones have to process many
thousands of documents per month. There are few ways to do this currently: (a)
manual data entry and processing, (b) template-based extraction model, or (c)
template-less machine learning approach. Manual data entry is not only
time-consuming and expensive but very error-prone as well. Template-based
approach requires an initial setup of hard-coded rules for every template, but
it still fails badly when an unseen template is encountered
\cite{str-template}. Template-less machine learning method tries to learn
generic features of various fields to extract so that they work well in
various templates but they need to be trained with a large number of
annotated documents to perform well.

In this paper, we propose a novel one-shot template matching algorithm that
brings the best of both worlds---template-based engine and template-less machine
learning. Our algorithm doesn't require any initial template setup, nor does it
need a very large amount of data to get high accuracy. Once provided with one
annotated document, future documents in the same format are processed
automatically with 90\% accuracy. We exploit the fact that for a specific
vendor and document type, the document format is very similar, i.e., the
position of annotated values and the neighboring keywords don't change much.
Moreover, if it extracts a field incorrectly, the user can correct it very
easily using our convenient review tool and subsequent documents in that format
will learn the corrections as well.

Our algorithm saves the contextual features of every annotated value that
includes information about both visual as well as textual features of not just
the actual value but the surrounding keywords as well, which is explained in
detail in Section~\ref{sec:method}. Our algorithm also automatically finds out whether a new
document belongs to any of the previously saved formats. To match a new
document with a saved template, we use a combination of image similarity \cite{svd}
and textual similarity \cite{levenshtein} metrics.

The rest of the paper is organized into four sections. In Section~\ref{sec:related}, we revisit the various previous approaches to solve
similar problems, and also mention how our approach stands out. In Section~\ref{sec:method}, we explain our algorithm in detail. We discuss the
experiments and their results in Section~\ref{sec:result}. Finally, in
Section~\ref{sec:conclusion}, we provide concluding remarks and possible
future works.
\section{Related Work}\label{sec:related}

Deciding whether two documents are of the same format requires a combination of image similarity and text similarity metrics. A number of perceptual hashing methods\cite{phash-overview,min-hash} have been used to detect near-duplicate images. Zeng et al. used eigenvalue matrix computed by Singular Value Decomposition (SVD) of image as features and computed similarity by comparing the angle between the eigenvalue matrices mapped into vector space \cite{svd}. Similarly, the most common approach for measuring textual similarity is the Levenshtein edit distance \cite{levenshtein}.

Flexible template-based extraction systems \cite{str-template,informys,apriori,intellix} locate the required text in the document by using the distance and direction from important surrounding keywords such as field labels. Cesarini et al.\cite{informys} only look at the nearest keyword whereas d'Andecy et al.\cite{apriori} computes distances and angles from every other word in the document and predicts the final location by averaging over the distances and angles from all words weighted by their \emph{itf-df} scores.

The first serious attempt at solving the automatic data capture problem using machine learning was made by Rossum\cite{rossum,rossum-table}. Trying to mimic human brain, they process documents in three stages: skim-reading, data localization, and precise reading. Holt et al.\cite{sypht} and Palm et al.\cite{cloudscan} used a content-based template-less machine learning approach that can classify any text block into one of predefined labels thereby claiming to work in unseen document formats as well. In \cite{sypht}, the authors reported 92.8\% percent accuracy after training the model with 300,000 documents. Completely vision object-detection models such as Faster-RCNN\cite{frcnn} and YOLO\cite{yolo,yolov3}, being trained on natural scenes, produce mixed results on document images. All these approaches require a large volume of annotated documents to train well.

Our method automatically utilizes template-features without needing any template-based rules. Since existing methods are either rigidly template-dependent or template-less, we cannot compare our work directly with any of them.

\section{Methodology}\label{sec:method}

Fig~\ref{fig:architecture} shows the high-level architecture of our model. There are three major steps: template matching, region proposal, and final area selection. These are explained in detail shortly. Our model maintains a database of unique annotated templates, takes a new document as input, and predicts the annotation for the new document if such a template exists in our database.

\begin{figure}[ht]
    \centering
    \includegraphics[width=0.9\linewidth]{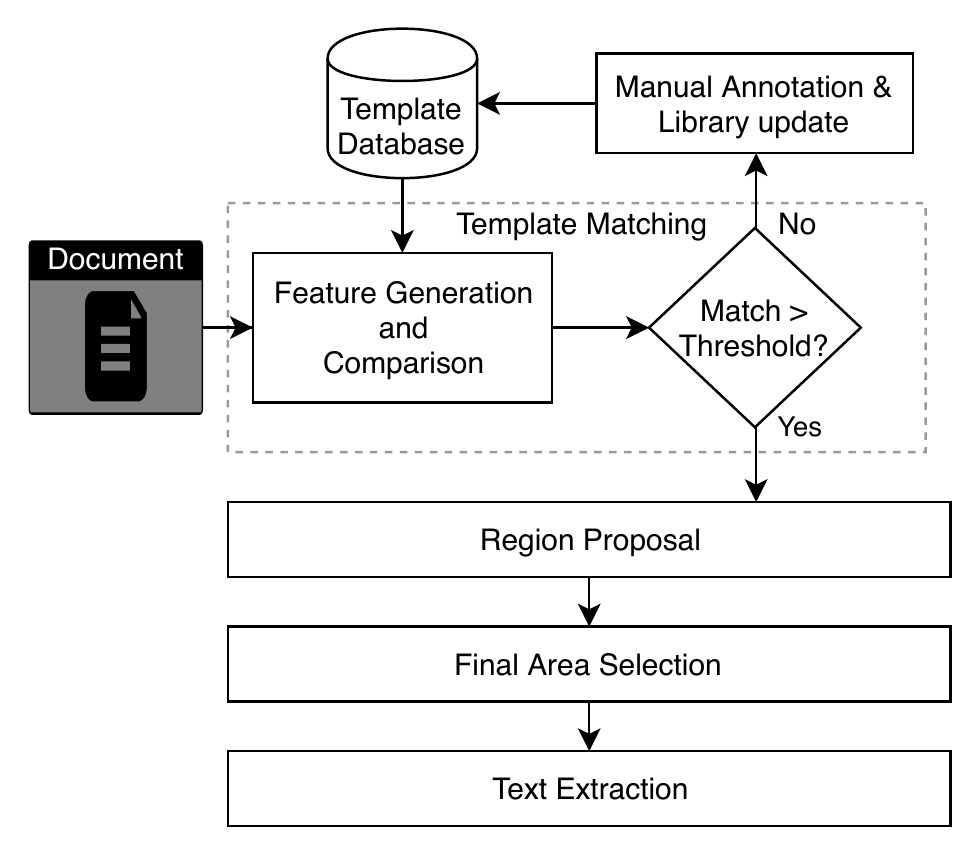}
    \caption{Model architecture (placeholder image)}
    \label{fig:architecture}
\end{figure}

\subsection{Optical Character Recognition (OCR)}

A new document may be an image, without any texts embedded. In that case, we can get the texts present in the document by OCR. For this research, OCR can be thought of as a black box that takes an image as input and outputs a list of words and their positions (bounding boxes). There are many high-quality commercial OCR engines such as Google Vision\cite{gvision}, Microsoft Azure Vision\cite{azure}, Amazon Textract\cite{aws-textract}. We used Amazon Textract for this research because they produced the best result as they were trained exclusively for document images.

\subsection{Template Matching}

In this step, a matching template from the database is chosen for the input document. If there is no match, the algorithm halts and the document is sent for manual annotation. We use a combination of visual and textual similarity measures to find a match. For image similarity, we compute the SVD of images and measure the cosine similarity between the $\Sigma$ diagonal matrices\cite{svd}. Equation \eqref{eq:svd} shows the SVD of a matrix $I$. However, before SVD, we perform some image preprocessing to adapt this metric to document images and make it invariant to image dimensions and lighting conditions.
%First, we resize the images for faster processing. Then we detect and extract edges using Canny edge detection\cite{canny} for lighting condition invariance. We then blur the image slightly to remove high-frequency noise and finally convert it to grayscale (monochrome).

Let $I$ = preprocessed input document image, and $T$ = preprocessed template document image.
\begin{align}
    (U_I, \Sigma_I, V_I) &= \text{SVD}(I) \label{eq:svd}\\
    (U_T, \Sigma_T, V_T) &= \text{SVD}(T) \nonumber
\end{align}
Now, the visual similarity $\text{Sim}_{\textit{visual}}$ is given by:
\begin{align}
    \text{Sim}_{\textit{visual}} = \cos\theta
    = \frac{\Sigma_I\cdot\Sigma_T}{\abs{\Sigma_I}\abs{\Sigma_T}}
    \in [-1,1]
\end{align}

For text similarity, we compute the fuzzy match\cite{fuzzywuzzy} (based on Levenshtein distance\cite{levenshtein}) of the top-$n$ and bottom-$n$ lines of text in the documents. Again, before the fuzzy matching, we perform some preprocessing steps to normalize the text.
\begin{align}
    \text{Sim}_{\textit{text}} = \textsc{Fuzzy-Match}(t_I, t_T) \in [0,1]
\end{align}
where, $t_I$ is the concatenation of the preprocessed top-$n$ and bottom-$n$ lines of text in the input document, and $t_T$ is the same for the template document. The combined similarity is simply the sum of visual and textual similarities:
\begin{align}
    \text{Sim}_{\textit{combined}} = \text{Sim}_{\textit{text}}
    + \text{Sim}_{\textit{visual}}
\end{align}

The template having the highest $\text{Sim}_{\textit{combined}}$, with $\text{Sim}_{\textit{text}} \geq C$, is selected as the final template for the input image, where $C$ is a manually set threshold. If $\text{Sim}_{\textit{text}} < C$ for all templates in the database, then the document is manually annotated and added to the database as a new template.

\subsection{Region Proposal}

Once the template document is selected, we use the template image and annotation to predict the approximate regions of all the fields in the input document. The annotation object is a JSON with field names as keys and the associated values and positions (top-left and bottom-right coordinates of the values) as the values. An example of an annotation object is given below:

\begin{lstlisting}[language=Python,caption={An annotation sample. The coordinates are calculated from the top-left corner of the document.},captionpos=b]
{"invoice_no":
    {"position": [53, 671, 452, 702],
     "value": "INV1234"},
 "date":
    {"position": [50, 635, 312, 666],
     "value": "2019-08-24"},
 "seller":
    {"position": [259, 27, 464, 58],
     "value": "ABC Pvt. Ltd."},
 "buyer":
    {"position": [821, 445, 1153, 468], 
     "value": "Zinc Enterprises"},
 "total":
    {"position": [48, 553, 419, 577],
     "value": "1,234.56"}} 
\end{lstlisting}

In this illustration, the annotation is a JSON object where the keys are the fields to be captured and the values have the position information of the text as well as the actual text for the field value. The ``position'' parameter contains the top-left $(x_{min},y_{min})$ and bottom-right $(x_{max},y_{max})$ coordinates of the rectangle surrounding the text.

Once we have the annotation of the matching template image, the following algorithm is used to get approximate region proposal for each field.
We use the correlation coefficient $R$ between the input document image and template image \cite{opencv_library} to obtain the approximate region-proposal.

\begin{equation}
R(x,y)= \sum _{x',y'} \left(T'(x',y') \cdot I'(x+x',y+y')\right)
\label{eq_corr}   
\end{equation}
where,
\begin{align*}
T'(x',y')&=T(x',y') - \frac{\sum _{x'',y''} T(x'',y'')}{w \cdot h}
\end{align*}
\begin{multline*}
I'(x+x',y+y') = I(x+x',y+y') \\
    - \frac{\sum _{x'',y''} I(x+x'',y+y'')}{w \cdot h}
\end{multline*}

\begin{algorithm}
\caption{Region Proposal}\label{alg:region-proposal}

\textbf{Input:} 
\begin{itemize}
    \item Preprocessed input document image ($I$).
    \item Preprocessed template document image ($T$).
    \item Template document annotation ($A_{T}$).
\end{itemize}

\textbf{Output:} Region proposals for all the fields in the input document.\\
\textbf{Procedure:}

\begin{algorithmic}[1] 
\label{region_prop}

% \STATE $w_I$ $\leftarrow$ \textsc{Width}($I$) \COMMENT{Input Image Width} 
\STATE $w_T$ $\leftarrow$  \textsc{Width}($T$) \COMMENT{Template Image Width} 
% \STATE $h_I \leftarrow$ $1.414*w_C$ \COMMENT{Input Image Height}
\STATE $h_T \leftarrow$  $1.414*w_T$\footnotemark{} \COMMENT{Template Image Height} 
\FOR{each field in $A_{T}$}
\STATE Get rectangular area of the field.\\ $(x_{min},y_{min},x_{max},y_{max})$
\STATE Increase the area of the rectangle slightly in all directions.\footnotemark{}\\
    %  $x_{min} \leftarrow \max(0, x_{min}-0.2w_T)$\\
    %  $y_{min} \leftarrow \max(0, y_{min}-0.2h_T)$\\
    %  $x_{max} \leftarrow \min(w_T, x_{max}+0.05w_T)$\\
    %  $y_{min} \leftarrow \min(h_T, y_{max}+0.05h_T)$
\STATE Crop out the new rectangular area in the template image. 
% \STATE Image transformation in the current image and cropped area in the template image to account for difference in lighting and noise.
\STATE Find the area in the input image where the cropped area from template image is most likely to match using \eqref{eq_corr}.
\ENDFOR
\end{algorithmic}
\end{algorithm}
\addtocounter{footnote}{-2} %3=n
\stepcounter{footnote}\footnotetext{This is done to make the regions have the same aspect ratio to handle documents with different aspect ratios.}
\stepcounter{footnote}\footnotetext{We expand the area in order to include some keywords common in both the template and the input image. Those keywords will help us accurately pinpoint the location of field values in the input regions proposed.}
% {We expand more above and to the left of the annotation because most of the keywords or labels are present there. Those keywords will help us accurately pinpoint the location of field values in the input regions proposed.}

Algorithm~\ref{alg:region-proposal} presents the pseudocode of our region proposal algorithm.

% There are cases where the number of pages in the template document and current document are unequal. This causes the image height of template and current document image to be vastly different but the width remains same. While proposing regions for each field in the current document, we use the area surrounding that field in the template document. This area is determined by taking some percent of the total image dimensions. The difference in image heights in the template and current document has to be accounted for before cropping out the areas. This makes the shape of the crop areas consistent. We take image width as standard and scale it by $1.414$ to obtain image height. 1.414 is the ratio of height to width in standard A4 size.%$(210mm X 297mm)$.

% Most of the fields to be captured in financial documents are present in the key-value form. The key is usually present in the top or left of the value. This is the intuition behind increasing the crop area in the template image to $20\%$ in the top and left directions and $5\%$ in the bottom and right directions. This increases the chance of matching the correct key-words in the template and current document. 

\subsection{Final Area Selection}

Next, we pinpoint the location of the annotation values in the input document by finding common words---present in both the input region and the template region of the field---and projecting the distance and displacement from the template region to the input region. This method was first devised in \cite{str-template}, but they looked for common words in whole document. We only look for common keywords inside the proposed regions for computational efficiency. Algorithm~\ref{alg:fin-area} shows the pseudocode for this algorithm.

% At this point, we have the approximate area in which value of each of the fields to be extracted lies. To detect the final area, we use Algorithm \ref{alg:fin-area} inspired from \cite{str-template}.

\begin{algorithm}
\caption{Final Area Selection}\label{alg:fin-area}
\textbf{Input:} 
\begin{itemize}
    \item Input document OCR
    \item Template document OCR
    \item Region Proposals $RP_I$ in input document (from Algorithm~\ref{alg:region-proposal})
    \item Input document and Template document dimensions.
\end{itemize}

\textbf{Output:} Final bounding-boxes for all fields in the input document.\\
\textbf{Procedure:}
\begin{algorithmic}[1] 
% \REQUIRE 
\FOR{each field in $RP_I$}
\STATE Find the texts that are common in this proposed area in the input image and corresponding area in the template image.
\IF{matches $>$ 0}
\STATE Find the text that is closest to the actual value for the field in the template image.
\STATE Get the vector from the center of the closest text and the actual value for the field in template image.
\STATE Normalize the vector with the dimensions of template image. De-normalize using the input image dimensions.
\STATE Using the vector in the input image, predict the center where the value of the field is present.
\STATE Using this center coordinates and the dimensions of rectangle surrounding the value of the field in the template image, obtain the rectangle in the input image using appropriate scaling.
\ELSE
\STATE Using the dimensions of rectangle surrounding the value of the field in the template image, obtain a rectangle at the center of the approximate area.
\ENDIF
% \STATE Get the text inside the rectangle predicted using input document OCR.
\ENDFOR
\end{algorithmic}
\end{algorithm}

\subsection{Text extraction}

Finally, once we have the final bounding box of the value, we  can extract the text from the OCR data. We extract all the words in the OCR data whose area overlaps with the proposed bounding box by more than a preset threshold and then combine them in natural reading order to get the final text for each of the fields.

% We extract all the words in the OCR data whose area overlaps with the proposed bounding box by more than $50\%$ and then combine them in proper reading order (left-to-right, top-to-bottom) based on their position.
\section{Experiment and Result}\label{sec:result}

\subsection{Dataset}

There are no publicly available dataset of modern business documents such as invoices, bank statements or employee forms, which is understandable given their strict confidentiality. Therefore, for this research, we acquired a dataset of 595 annotated invoices from a large international invoice financing company. All of them were in the English language and there were about 35 unique formats or templates. For fields to extract, we considered the most common ones: (a) invoice number, (b) date, (c) seller, (d) buyer, and (e) total due. We used one sample each of every template as the training set and the rest 560 documents as the test set. Again, due to confidential reasons, we cannot make the dataset public.

\subsection{Evaluation Metrics}

The ground truth values don't have positional values, so we can't compute the quality of output bounding boxes. Therefore, we evaluate our model by comparing the output values of the extracted fields. Since text output may have few erroneous characters, mostly due to OCR error, we define two metrics for evaluation---\textsc{Mean-Fuzzy-Match} and \textsc{Accuracy}---as follows:

\begin{align}
  \textsc{Mean-Fuzzy-Match} &=
  \frac{ \sum_{i=1}^{N} \textsc{Fuzzy-Match}(\hat{y_i}, y_i) }{N}
\end{align}
\begin{align}
  \textsc{Accuracy} &= \frac{
    % \sum_{i,\hat{y}_i=y_i} 1
    \text{no. of samples where } \hat{y}_i = y_i
  }{N}
\end{align}
where, $\hat{y}$ and $y$ are the output and ground truth values respectively both with length $N$, and \textsc{Fuzzy-Match} $\in [0, 1]$ is the fuzzy text matching function based on Levenshtein distance. In our implementation, we used the \texttt{fuzzywuzzy} library\cite{fuzzywuzzy} for this.

The \textsc{Accuracy}, which checks for exact match between the predicted value and the ground-truth, is affected by minor OCR errors (such as recognizing ``0'' as ``O''). We include the \textsc{Mean-Fuzzy-Match} metric to see how our model would perform in cases where exact match isn't required.

\subsection{Result}

The results of our model on the 560 test invoices are shown in Table~\ref{tab:result}. We can see that \textsc{Mean-Fuzzy-Match} is significantly greater than \textsc{Accuracy}, implying that our model can leverage better accuracy if the minor OCR errors are corrected by post-processing. For instance, the buyer and seller names can be matched with a lookup table. Similarly, dates and amounts can be canonicalized to eliminate format discrepancies. Other texts can be normalized by trimming whitespaces, converting to lowercase, and so on.

\begin{table}[ht]
    \caption{The performance of our model.}
    \label{tab:result}
    \centering
    \begin{tabular}{lcc}
        \toprule
        \textbf{Field} & \textbf{\textsc{Accuracy}} & \textbf{\textsc{Mean-Fuzzy-Match}}\\
        \midrule
        Invoice number~~~~~~~~ & 79.2 & 80.7\\
        Date & 86.4 & 89.4 \\
        Seller & 91.5 & 93.8 \\
        Buyer & 90.2 & 94.1 \\
        Total due & 84.7 & 88.2 \\
        \midrule
        \textbf{Overall} & \textbf{86.4} & \textbf{89.2}\\
        \bottomrule
    \end{tabular}
\end{table}

Considering the fact that our model doesn't require per-template rules and requires very few training samples, combined with our easy review tool, getting over 86\% accuracy can result in a significant reduction in time, cost, and effort it takes for businesses to process documents.

% \subsection{Time Complexity}

% Can we write something about time complexity so that we can mention it as an area of future improvement?
\section{Conclusion and Future Work}\label{sec:conclusion}

In this paper, we presented a new way of solving the problem of automatic data capture from documents. Requiring only one example per template, our method is very effective for dataset with recurring templates.

This research has many areas for improvement though. First of all, it can't handle multi-page documents. Future works can attempt to tackle this. In addition, our model, which right now only looks at only one saved sample to predict the outputs, can be made to predict based on all saved samples of the specific template to generalize better and improve overall accuracy. Also, further research can be done to make it work with recurring fields like table line items.

\section*{Acknowledgment}

We would like to thank Rushabh Sheth for providing us the required funding and resources. We are also grateful for the Documso annotators for manually labeling the dataset.

% \printbibliography
\bibliographystyle{IEEEtran}
\urlstyle{tt}
\bibliography{main}

\end{document}